\documentclass[11pt]{article}
\usepackage{tikz}
\usetikzlibrary{math,shapes}
\usetikzlibrary{shapes.geometric,positioning,}

\usepackage{xr}
\usepackage{hyperref}

\usepackage[margin=1in]{geometry}
\usepackage{amsmath, amssymb, amsthm}
\usepackage{booktabs}
\usepackage{natbib}
\usepackage{enumerate}
\usepackage{centernot}
\usepackage{authblk}

\graphicspath{{./}{figures/}}

\newtheorem{definition}{Definition}

\begin{document}

% \journaltitle{American Journal of Epidemiology}
% \DOI{DOI HERE}
% \copyrightyear{2022}
% \pubyear{2023}
% \access{Advance Access Publication Date: Day Month Year}
% \appnotes{Paper}

% \firstpage{1}

%\subtitle{Subject Section}

\title{Counterfactual Harm: A Counter-argument}

\author{Amit N. Sawant\thanks{Corresponding author: amit.sawant@epfl.ch}, Mats J. Stensrud}
% \author[3]{Third Author}
% \author[3]{Fourth Author}
% \author[4]{Fifth Author\ORCID{0000-0000-0000-0000}}

\affil{Institute of Mathematics, \'Ecole Polytechnique F\'ed\'erale de Lausanne, Switzerland}

% \email{amit.sawant@epfl.ch}
% \authormark{Sawant et al.}

% \address[1]{\orgdiv{Chair of Biostatistics, Institute of Mathematics}, \orgname{Ecole Polytechnique Fédérale de Lausanne}, \orgaddress{\street{Lausanne}, \postcode{1015}, \state{VD}, \country{Switzerland}}}
% \address[2]{\orgdiv{Department}, \orgname{Organization}, \orgaddress{\street{Street}, \postcode{Postcode}, \state{State}, \country{Country}}}
% \address[3]{\orgdiv{Department}, \orgname{Organization}, \orgaddress{\street{Street}, \postcode{Postcode}, \state{State}, \country{Country}}}
% \address[4]{\orgdiv{Department}, \orgname{Organization}, \orgaddress{\street{Street}, \postcode{Postcode}, \state{State}, \country{Country}}}

% \corresp[$\ast$]{Corresponding author. \href{email:amit.sawant@epfl.ch}{amit.sawant@epfl.ch}}

% \received{Date}{0}{Year}
% \revised{Date}{0}{Year}
% \accepted{Date}{0}{Year}
\maketitle

\begin{abstract}
    As AI systems are increasingly used to guide decisions, it is essential that they follow ethical principles. A core principle in medicine is non-maleficence, often equated with ``do no harm''. A formal definition of harm based on counterfactual reasoning has been proposed and popularized. This notion of harm has been promoted in simple settings with binary treatments and outcomes. Here, we highlight a problem with this definition in settings involving multiple treatment options. Illustrated by an example with three tuberculosis treatments (say, A, B, and C), we demonstrate that the counterfactual definition of harm can produce intransitive results: B is less harmful than A, C is less harmful than B, yet C is more harmful than A when compared pairwise. This intransitivity poses a challenge as it may lead to practical (clinical) decisions that are difficult to justify or defend. In contrast, an interventionist definition of harm based on expected utility forgoes counterfactual comparisons and ensures transitive treatment rankings.
    
\end{abstract}

% \keywords{}

\section{Introduction}
\label{sec:intro}

The Hippocratic maxim of ``First, do no harm'' remains a principle in modern medicine. However, the question of what constitutes harm in a medical setting is unclear and has led to rich debate. Unambiguous definitions of harm are important to assess the performance of algorithms, which are increasingly being used for decision-making. For instance, \citet{andrikyan2024artificial} stated that a considerable portion of medical advice given by artificial intelligence (AI)-powered chatbots was deemed `potentially harmful' by medical experts. A concrete definition of harm helps ensure that algorithm-based decision making also follows the principle of ``doing no harm''.

Characterizing harm in mathematical terms has been variously attempted, see e.g., \citet{sarvet2023perspective} for a review. In particular, \citet{kallus2022s, richens2022counterfactual, mueller2023personalized, ben2024policy} considered a counterfactual notion of harm, which requires comparison of joint (cross-world) counterfactuals. \citet{sarvet2023perspective} also discussed an interventionist definition of harm, which is defined with respect to single-world quantities \citep{richardson2013single}. All of these works, however, were considering the setting where treatments are binary. % Some other articles did consider the possibility of more than two treatment options when defining harm in counterfactual terms \citep{beckers2022quantifying, beckers2024causal, richens2022counterfactual}. 

Most existing work on counterfactual harm, but not interventionist harm, is further restricted to binary outcomes, e.g., \citet{kallus2022s, mueller2023personalized, ben2024policy, gelman2025russian}. With binary outcomes, \citet{dawid2023personalised, sarvet2023perspective, sarvet2024rejoinder} discussed the interventionist philosophy and clarified that certain notions of harm, which were claimed to be impossible to express in interventionist terms \citep{mueller2023personalized,mueller2023perspective}, have a clear interventionist formulation. Similarly, \citet{kallus2022s} mentioned that the key criterion for determining harm was simply that the average treatment effect was negative, thereby reducing to an interventionist criterion. 

For many medical conditions, there are more than two treatment options. Also, outcomes of interest are not as simple as dichotomies. Diseases can get progressively worse without the patient dying. Patients may discontinue treatment due to unforeseen side-effects without being fully cured. Patients who initially received no treatment may be put on rescue treatment, without them dying. %Considering all such outcomes and mapping them to a binary scale is reductive.

With non-binary outcomes and singular treatments, results from the counterfactual definition of harm tend to be uninformative under plausible assumptions \citep{cui2023policy, fan2010sharp}; it is often not possible to determine from a counterfactual perspective whether a treatment is more beneficial than harmful \citep{zhang2024identifying, fava2024predicting, de2025probability, gechter2024generalizing} without making strong, and in-principle untestable, cross-world assumptions \citep{richardson2013single}.

%One could hope that the arguments from the binary setting would translate to the setting with more than two treatments and outcomes. However, 

Regardless of whether counterfactual definitions of harm result in informative comparisons, subtle problems could arise. Illustrated by an example on tuberculosis treatment, we show that when we have more than one treatment, comparing treatments pairwise using the counterfactual definition of harm can lead to intransitivity in the total order of treatments. That is, if our objective is to minimize counterfactual harm, then a treatment $A=a_{2}$ is better than the baseline treatment $A=a_{1}$. Treatment $A=a_{3}$ is then better than treatment $A=a_{2}$ by the same criterion, but treatment $A=a_{3}$ is worse than treatment $A=a_{1}$ in a direct comparison, which is problematic. On the other hand, using the criterion of interventionist harm does not lead to intransitivity. Transitivity is often desirable in many settings. For example, transitivity is an axiom in the von Neumann–Morgenstern utility theorem \citep{von1947theory}.

We start with basic definitions of counterfactual and interventionist harm, as discussed in detail in existing works, e.g., \citep{richens2022counterfactual, ben2024policy, sarvet2023perspective}. % is deferred to Appendix \ref{app:other papers}.

\section{Notions of harm}
\label{sec:definition of harm}

Consider a treatment $A$ that takes binary values $(A\in \{0, 1\})$ and an outcome of interest $Y$. For the sake of illustration, we will use discrete, ordinal outcomes, but our example can be readily adapted to continuous real-valued outcomes. 

It is clear that some outcomes are always better than others. To be explicit, consider three outcomes as $y_{1}:$ death, $y_{2}:$ discontinuing treatment, and $y_{3}:$ being fully cured. Discontinuing treatment is preferable to death, even though it is a relatively worse outcome to being cured, meaning the outcomes can be ordered as $y_{1} < y_{2} < y_{3}$. Note, we overload the comparison operator `$<$' to both compare one event or outcome to another, along with the conventional sense of comparing two numbers.
% For example, it is better to be fully cured of a disease than it is to discontinue treatment without being cured. Either of these outcomes are better than dying of the disease. Thus, outcomes $Y$ are assigned ranks $(r) = Y)$ for a ranking function $\rho$. 

We can additionally define a utility function $\mu$ that assigns a specific real-valued utility to each outcome $Y$. The utility function is order-preserving so that if $y_{1} \leq y_{2}$, then $\mu(y_{1}) \leq \mu(y_{2})$. Without loss of generality, we assume that outcomes with a greater numerical utility are better.
% The ranking function $\rho$ is order-preserving. 

For any individual $i$, there exist two counterfactual variables $Y_{i}^{a=1}$ and $Y_{i}^{a=0}$, which are respectively the potential outcomes if the individual were to be administered the treatment $(a=1)$ or not $(a=0)$. We omit the subscript $i$ where the interpretation is obvious, for brevity in notation.

We first describe the counterfactual definition of harm, which is commonly used in the literature \citep{kallus2022s, richens2022counterfactual, mueller2023personalized, ben2024policy, fava2024predicting}. If the potential outcome under treatment is worse than the potential outcome under no treatment, then the treatment is counterfactually harmful for that individual. This can be expressed concisely using indicator functions as

\begin{equation}
\label{eq:harm definition}
    \text{harm}_{i} = I(Y_{i}^{a=1} < Y_{i}^{a=0}).
\end{equation}

Analogously, we can define the treatment to be beneficial for individual $i$ if 
\begin{equation}
\label{eq:benefit definition}
    \text{benefit}_{i} = I(Y_{i}^{a=1} > Y_{i}^{a=0}).
\end{equation}

Now suppose there is more than one treatment option. Let $\mathcal{A} = \{a_{1}, a_{2},\dots a_{n}\}$ be the set of all treatments available, including the possibility of no treatment. %The ranks span an ordinal scale from $y_{1}$ to $(m)$. There exists a utility function $\mu(Y)$ mapping outcomes to a real-valued utility. 
When there is more than one treatment, we define ``doing no harm'' against the existing standard-of-care. Suppose a treatment $A=a_{1}$ is the current standard-of-care. Another treatment $A=a_{2}$ is considered harmful for an individual if $Y^{a_{2}} < Y^{a_{1}}$. It is equivalent to say the treatment is harmful if $\mu(Y^{a_{2}}) < \mu(Y^{a_{1}})$. %Doing benefit can be defined in an isomorphic way (Equations \ref{eq:harm definition}, \ref{eq:benefit definition}), but we restrict our attention to harm in the main text.  

% BLAH BLAH BLAH

It is impossible to observe more than one potential outcome $Y_{i}^{a}$ for any individual. Thus, evidence of counterfactual harm can only be identified at a population level, say, as $P(Y^{a_{2}} < Y^{a_{1}}) = \mathbb{E}[I(Y^{a_{2}} < Y^{a_{1}})]$, which is the probability of harm in the population. 

It may be tempting to simplify the expression $P(Y^{a_{2}} < Y^{a_{1}})$ to $P(Y^{a_{2}} - Y^{a_{1}} < 0)$ but we should remember that outcomes $Y^{a}$ are defined to be events from an ordered set $\mathcal{Y}$. The sum or difference of two such $y$'s is ambiguous. In terms of utility however, we can indeed make sense of the statement $\mu(Y^{a_{2}}) - \mu(Y^{a_{1}}) < 0$ and consider its expectation over a population, since utilities are real-valued. Thus, we get an alternative definition of harm in terms of utility. A treatment $A=a_{2}$ is harmful compared to the standard-of-care $A=a_{1}$ if $\mathbb{E}[\mu(Y^{a_{2}}) - \mu(Y^{a_{1}})] = \mathbb{E}[\mu(Y^{a_{2}})] - \mathbb{E}[\mu(Y^{a_{1}})] < 0$.

Defining harm in terms of utility at the population level as opposed to individual outcomes has been termed as `interventionist harm' by \citet{sarvet2023perspective}. We thus have two different definitions of harm as follows:

\begin{definition}[Counterfactual harm]
\label{def:counterfactual harm}
    For a treatment $A=a_{2}$, counterfactual harm exists if $\mathbb{E}[I(Y^{a_{2}} < Y^{a_{1}})] > 0$, where $A=a_{1}$ is the existing standard-of-care.
\end{definition}

\begin{definition}[Interventionist harm]
\label{def:interventionist harm}
    For a treatment $A=a_{2}$, interventionist harm exists if $\mathbb{E}[\mu(Y^{a_{2}})] -\mathbb{E}[\mu(Y^{a_{1}})] < 0$, where $A=a_{1}$ is the existing standard-of-care.
\end{definition}

To assess if a treatment is harmful in the population, by either definition, we require functionals of the counterfactual distribution to be estimable. Specifically, consider a Randomized Control Trial (RCT) with two arms, the new treatment $A=a_{2}$ and the existing standard-of-care $A=a_{1}$. The marginal distribution of each counterfactual is identified in this trial, but the joint counterfactual distribution of $(Y^{a_{2}}, Y^{a_{1}})$ remains unidentified.

Interventionist harm is point-identified whenever individual treatment effects are point-identified and does not require consideration of the joint counterfactual distribution \citep{stensrud2024optimal}. On the other hand, we can use the marginal distributions to derive bounds for the probability of counterfactual harm. We now consider a simple situation with multiple treatment options for tuberculosis, and we will compare treatments using the two definitions of harm (Definitions \ref{def:counterfactual harm}, \ref{def:interventionist harm}). 

\section{An example on tuberculosis and intransitivity}
\label{sec:paradox}

Tuberculosis (TB) is an infectious disease that is prevalent particularly in South-East Asia. Most patients with a TB diagnosis have latent TB, which can progress to active disease if left untreated and even result in death \citep{world2008implementing}.

Effective TB treatment requires long-lasting medication courses of up to eight months with a mixture of antibiotics \citep{world2008implementing}. There are strong antibiotics which can cause side-effects such as nausea and vomiting \citep{world2010treatment}, and also weak antibiotics. If there is incomplete eradication of the TB pathogen, due to inadequate compliance with the treatment or ineffective antibiotics, the patient develops drug-resistant TB which has a higher risk of progressing to active TB and death compared to drug-susceptible TB.

Thus, there is genuine concern that TB medication may be harmful. We describe a hypothetical scenario where pair-wise comparison of treatments to minimize harm gives an intransitive overall ordering. This example has been adapted from results on intransitive orderings from \citet{blyth1972some} and \citet{pasciuto2016mystery, grime2017bizarre}

We consider outcomes $\mathcal{Y} = \{y_{1}, y_{2}, \ldots y_{6}\}$ that can be uncontroversially ordered as follows:

\begin{enumerate}[$y_1:$]
    \item Death from TB within one year.
    \item Extensively drug-resistant (XDR)-TB, resistant to strong antibiotics.
    \item Multiple drug-resistant (MDR)-TB, resistant to weak antibiotics.
    \item Latent TB.
    \item Fully cured of TB with some side-effects.
    \item Fully cured of TB without side-effects.
\end{enumerate}

We define the utility function $\mu_{1}(Y)$ that simply outputs the numerical position of the outcome in the ordered set $\mathcal{Y}$ as the utility. If a patient dies of TB within one year, the outcome $y_{1}$ has utility equal to $1$, as it is the first element in $\mathcal{Y}$. The utility function $\mu_{1}$ is chosen for illustrative purposes and can be replaced with any other order-preserving utility function $\mu$.

\subsection{Initial diagnosis}
\label{sec:no treatment}

Consider a patient with a diagnosis of latent TB. When patients with latent TB are left untreated $(A=a_{1})$, one-sixth of the patients develop active disease and die within one year. The remaining five-sixths still have latent TB which may progress to active disease later on. Thus, under the assumption of causal consistency \citep{hernan2024causal}, which should be uncontroversial in our example,
\begin{align*}
    P(Y^{a_{1}} = y_{1}) &= P(Y = y_{1} \mid A=a_{1}) = 1/6,\\
    P(Y^{a_{1}} = y_{4}) &= P(Y = y_{4} \mid A=a_{1}) = 5/6.
\end{align*}

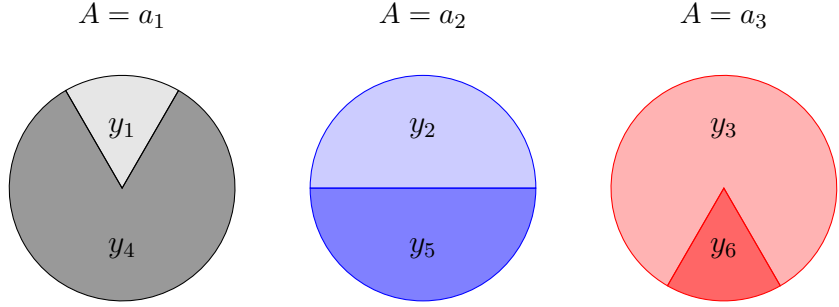
\begin{figure*}[htbp]
    \centering
    \begin{tikzpicture}
        \tikzmath{\r=1.5;\lab=0.8;\trt=2;}
        \coordinate (center) at (-4,0);
        \filldraw[draw=black,fill=black, fill opacity = 0.10] (center) -- +(60:\r) arc[start angle=60,end angle=120,radius=\r] -- cycle;
        \filldraw[draw=black,fill=black, fill opacity = 0.40] (center) -- +(120:\r) arc[start angle=120,end angle=420,radius=\r] -- cycle;
        \node[font=\large] at ([shift={(90:\lab)}]center) {$y_{1}$};
        \node[font=\large] at ([shift={(270:\lab)}]center) {$y_{4}$};
        \node [above= \trt of center] {$A=a_{1}$};
        
        \coordinate (center) at (0,0);
        \filldraw[draw=blue,fill=blue, fill opacity = 0.20] (center) -- +(0:\r) arc[start angle=0,end angle=180,radius=\r] -- cycle;
        \filldraw[draw=blue,fill=blue, fill opacity = 0.50] (center) -- +(180:\r) arc[start angle=180,end angle=360,radius=\r] -- cycle;
        \node[font=\large] at ([shift={(90:\lab)}]center) {$y_{2}$};
        \node[font=\large] at ([shift={(270:\lab)}]center) {$y_{5}$};
        \node [above= \trt of center] {$A=a_{2}$};
        
        \coordinate (center) at (4,0);
        \filldraw[draw=red,fill=red, fill opacity = 0.30] (center) -- +(-60:\r) arc[start angle=-60,end angle=240,radius=\r] -- cycle;
        \filldraw[draw=red,fill=red, fill opacity = 0.60] (center) -- +(240:\r) arc[start angle=240,end angle=300,radius=\r] -- cycle;
        \node[font=\large] at ([shift={(90:\lab)}]center) {$y_{3}$};
        \node[font=\large] at ([shift={(270:\lab)}]center) {$y_{6}$};
        \node [above= \trt of center] {$A=a_{3}$};
    \end{tikzpicture}
    \caption{Pie charts representing potential outcomes $P(Y^{a}=y_{r})$ under `No Treatment' $(A=a_{1})$, `Strong Antibiotics' $(A=a_{2})$, and `Weak Antibiotics' $(A=a_{3})$ respectively.}
    \label{fig:pie charts}
\end{figure*}

We can also depict this graphically as in Figure \ref{fig:pie charts}.

\subsection{Treatment with strong antibiotics}
\label{sec:rct 1}

In an RCT (RCT-1) that compared strong antibiotics administered for one year $(A=a_{2})$ to no treatment $(A=a_{1})$, it was found that half of the patients in the treatment arm were fully cured of TB, and remarkably no patients in the treatment arm died of active disease. However, all of the patients experienced nausea throughout the course of medication, due to the strong immune response elicited by the antibiotics. 

% there was no change in outcomes under no treatment. One-sixth of patients in the control arm still died by the end of the year. In contrast, 

Owing to the side-effects and the long duration of treatment, half of the patients in the treatment arm developed XDR-TB due to incomplete adherence. Outcomes in the control arm under no treatment were the same as before. The results of the RCT can be summarized as follows,
\begin{align*}
    P(Y^{a_{2}} = y_{2}) &= P(Y = y_{2} \mid A=a_{2}) = 1/2, \\
    P(Y^{a_{2}} = y_{5}) &= P(Y = y_{5} \mid A=a_{2}) = 1/2, \\
    P(Y^{a_{1}} = y_{1}) &= P(Y = y_{1} \mid A=a_{1}) = 1/6, \\
    P(Y^{a_{1}} = y_{4}) &= P(Y = y_{4} \mid A=a_{1}) = 5/6.
\end{align*}

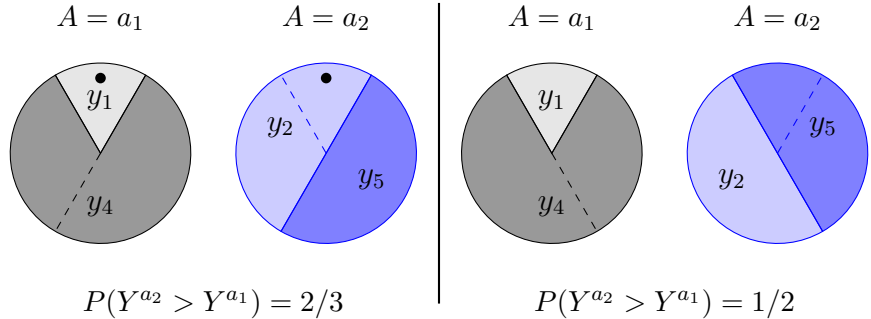
\begin{figure*}[htbp]
    \centering
    \begin{tikzpicture}
        \tikzmath{\r=1.2;\lab=0.7;\trt=1.5;}
        \coordinate (center) at (-4.5, 0);
        \filldraw[draw=black,fill=black, fill opacity = 0.10] (center) -- +(60:\r) arc[start angle=60,end angle=120,radius=\r] -- cycle;
        \filldraw[draw=black,fill=black, fill opacity = 0.40] (center) -- +(120:\r) arc[start angle=120,end angle=420,radius=\r] -- cycle;
        \node[font=\large] at ([shift={(90:\lab)}]center) {$y_{1}$};
        \node[font=\large] at ([shift={(270:\lab)}]center) {$y_{4}$};
        \fill[fill=black] ([shift={(90:\lab+0.3)}]center)  circle[radius=2pt];
        \node [above= \trt of center] {$A=a_{1}$};
        \draw[black, dashed] (center) -- +(240:\r);
        \coordinate (center) at (-1.5, 0);
        \filldraw[draw=blue,fill=blue, fill opacity = 0.20] (center) -- +(60:\r) arc[start angle=60,end angle=240,radius=\r] -- cycle;
        \filldraw[draw=blue,fill=blue, fill opacity = 0.50] (center) -- +(240:\r) arc[start angle=240,end angle=420,radius=\r] -- cycle;
        \node[font=\large] at ([shift={(150:\lab)}]center) {$y_{2}$};
        \node[font=\large] at ([shift={(330:\lab)}]center) {$y_{5}$};
        \fill[fill=black] ([shift={(90:\lab+0.3)}]center)  circle[radius=2pt];
        \draw[blue, dashed] (center) -- +(120:\r);
       
        \node [above= \trt of center] {$A=a_{2}$};
        \node at (-3, -2) {$P(Y^{a_{2}} > Y^{a_{1}}) = 2/3$};

        \draw[black, thick] (0, 2) -- (0, -2);

        \coordinate (center) at (1.5, 0);
        \filldraw[draw=black,fill=black, fill opacity = 0.10] (center) -- +(60:\r) arc[start angle=60,end angle=120,radius=\r] -- cycle;
        \filldraw[draw=black,fill=black, fill opacity = 0.40] (center) -- +(120:\r) arc[start angle=120,end angle=420,radius=\r] -- cycle;
        \node[font=\large] at ([shift={(90:\lab)}]center) {$y_{1}$};
        \node[font=\large] at ([shift={(270:\lab)}]center) {$y_{4}$};
        % \fill[fill=black] ([shift={(90:\lab+0.3)}]center)  circle[radius=2pt];
        \node [above= \trt of center] {$A=a_{1}$};
        \draw[black, dashed] (center) -- +(-60:\r);
        \coordinate (center) at (4.5, 0);
        \filldraw[draw=blue,fill=blue, fill opacity = 0.20] (center) -- +(120:\r) arc[start angle=120,end angle=300,radius=\r] -- cycle;
        \filldraw[draw=blue,fill=blue, fill opacity = 0.50] (center) -- +(-60:\r) arc[start angle=-60,end angle=120,radius=\r] -- cycle;
        \node[font=\large] at ([shift={(210:\lab)}]center) {$y_{2}$};
        \node[font=\large] at ([shift={(30:\lab)}]center) {$y_{5}$};
        % \fill[fill=black] ([shift={(90:\lab+0.3)}]center)  circle[radius=2pt];
        \draw[blue, dashed] (center) -- +(60:\r);
        
        \node [above= \trt of center] {$A=a_{2}$};
        \node at (3, -2) {$P(Y^{a_{2}} > Y^{a_{1}}) = 1/2$};
    \end{tikzpicture}
    \caption{Joint potential outcomes for patients in RCT-1. Consider first the left set of pies, which shows a joint distribution of $(Y^{a_{1}}, Y^{a_{2}})$ that is compatible with RCT-1 and gives the maximum proportion of patients benefiting from treatment $A=a_{2}$ over $A=a_{1}$. The dot `\textbullet' indicates a patient with the two potential outcomes $Y^{a_{1}} = y_{1}, Y^{a_{2}} = y_{2}$, and it is understood that $P(Y^{a_{1}} = y_{1}, Y^{a_{2}} = y_{2}) = 1/6$. The right set of pies is also compatible with RCT-1, and gives the minimum proportion of patients benefiting from treatment $A=a_{2}$ over $A=a_{1}$; in particular $P(Y^{a_{1}} = y_{1}, Y^{a_{2}} = y_{2}) = 0$. Dashed lines are intended as guides for visual comparison and do not indicate actual partitions of the data.}
    \label{fig:rct 1}
\end{figure*}

Figure \ref{fig:rct 1} shows the results of RCT-1 in graphical form. We can compute the probability of benefit and harm following Definition \ref{def:counterfactual harm} based on the data from RCT-1. The procedure described here follows from classical Fr\'echet-Hoeffding bounds \citep{frechet1935generalisation, ruschendorf1991frechet} widely used in the literature on counterfactual harm \citep{fan2010sharp, kallus2022s, de2025probability}. A short mathematical argument replicating the same is given in Appendix \ref{app:frechet bounds}.

The left set of Figure \ref{fig:rct 1} shows one possibility of the joint distribution of the two potential outcomes. All the patients who would have died under no treatment, would have survived with XDR-TB if given strong antibiotics treatment. Thus, this one-sixth of the population benefited from taking treatment $A=a_{2}$. In addition, half of the patients in the control arm, who would have survived with latent TB under no treatment, would have been fully cured with the antibiotics. Therefore, the maximum proportion of patients that would have benefited from the treatment is $1/6+1/2=2/3$.

The right set of Figure \ref{fig:rct 1} represents the other extreme of the joint distribution. All the patients who would have died under no treatment, would be fully cured under treatment. In addition, one-third of the entire population, who would have survived with latent TB under no treatment, would have been fully cured with the antibiotics. Therefore, the minimum proportion of patients that would have benefited from the treatment is $1/2$.

In RCT-1 $P(\text{benefit}) = P(Y^{a_{2}} > Y^{a_{1}})$ is therefore bounded,

\begin{equation}
    1/2\ \leq\ P(Y^{a_{2}} > Y^{a_{1}})\ \leq \ 2/3. 
\end{equation}

With probability greater than or equal to half, the patient benefits from treatment $A=a_{2}$ over no treatment. Correspondingly, with at most probability $1/2$, the patient would be harmed by taking strong antibiotics. In addition, the edge case, $P(\text{benefit}) = P(\text{harm}) = 1/2$, occurs only if joint outcomes are strongly negatively correlated, i.e. any patient with $Y^{a_{1}} = y_{1}$ always has $Y^{a_{2}} = y_{5}$. If even one patient in the population has the joint outcome $(Y^{a_{1}} = y_{1}, Y^{a_{2}} = y_{2})$, then the proportion of patients benefiting from strong antibiotics is strictly greater than $1/2$, see Appendix \ref{app:frechet bounds} for a rigorous proof.
% at least half of patients benefit from treatment $A=a_{2}$ over no treatment. Correspondingly, at most half of patients benefit from taking no treatment over treatment $A=a_{2}$.

From a perspective of maximizing benefit/minimizing harm in the counterfactual sense, it is clear that prescribing antibiotics is better than giving no treatment. We employ the simple decision rule that we switch to the new treatment if $P(\text{benefit}) \geq P(\text{harm})$. Later on in Section \ref{sec:discussion}, we consider decision rules that weight benefit and harm differently.

The expected utility under no treatment is $\mathbb{E}[\mu_{1}(Y^{a_{1}})] = 3.5$, and that under strong antibiotic treatment $\mathbb{E}[\mu_{1}(Y^{a_{2}})]$ is also $3.5$. The loss in utility for the patients who develop XDR-TB is offset by the gain in utility for the patients who are fully cured. Thus, there is no evidence of interventionist harm (or benefit) in this case.

Although there is a non-zero probability of counterfactual harm to the patient, it is impossible to know before prescribing medication whether the patient will be counterfactually harmed by treatment with antibiotics. Even if the patient has a worse outcome and would be considered ``harmed'' by this definition, the outcome of XDR-TB is better than dying, which is the worst outcome under no treatment. %\citet{mueller2023personalized} describe augmentation of experimental and observational data to identify the subset of the population that would be harmed, but it relies on determinisms in the data that may not be true in general.

We must take a call between prescribing no treatment and risking death of the patient, or prescribing antibiotics and risking XDR-TB. This is a tradeoff that exists in many practical settings. The utility function $\mu$ in the interventionist approach arguably serves to balance such tradeoffs. Some might argue that it is best to choose the treatment that avoids death as far as possible. To this end, we can enforce a utility function $\mu$ that maps ordered outcomes in $\mathcal{Y}$ to binary outcomes of death and survival. We consider utility functions other than $\mu_{1}$ in Section \ref{sec:discussion}. %On the whole, treatment with strong antibiotics provides an improvement in the standard-of-care.

% The proportion of population that would be harmed by withholding treatment is greater than the proportion that would be harmed by giving treatment

\subsection{Treatment with weak antibiotics}
\label{sec:rct 2}

Unpleasant side effects over the course of medication are undesirable. In another RCT (RCT-2), newer antibiotics $(A=a_{3})$ that did not cause side effects were compared to strong antibiotics $(A=a_{2})$ as the control.
% Suppose that, after some months, newer antibiotics are found to combat TB that do not cause side-effects. A new treatment regimen is assessed in another RCT (RCT-2), where patients are assigned the new antibiotics $(A=a_{3})$ or the old antibiotics $(A=a_{2})$, which is the existing standard-of-care. 

In RCT-2, one-sixth of patients in the treatment arm were fully cured within one year, with no side-effects. The new antibiotics did not elicit a strong immune response however, so five-sixths of the patients ended up with MDR-TB. The results from the control arm under treatment $A=a_{2}$ were the same as those in RCT-1. The results of RCT-2 can be summarized as follows,
\begin{align*}
    P(Y^{a_{3}} = y_{3}) &= P(Y = y_{3} \mid A=a_{3}) = 5/6, \\
    P(Y^{a_{3}} = y_{6}) &= P(Y = y_{6} \mid A=a_{3}) = 1/6, \\
    P(Y^{a_{2}} = y_{2}) &= P(Y = y_{2} \mid A=a_{2}) = 1/2, \\
    P(Y^{a_{2}} = y_{5}) &= P(Y = y_{5} \mid A=a_{2}) = 1/2.
\end{align*}

\begin{figure*}[htbp]
    \centering
    \begin{tikzpicture}
        \tikzmath{\r=1.2;\lab=0.7;\trt=1.5;}
        \coordinate (center) at (-4.5,0);
        \filldraw[draw=blue,fill=blue, fill opacity = 0.20] (center) -- +(0:\r) arc[start angle=0,end angle=180,radius=\r] -- cycle;
        \filldraw[draw=blue,fill=blue, fill opacity = 0.50] (center) -- +(180:\r) arc[start angle=180,end angle=360,radius=\r] -- cycle;
        \node[font=\large] at ([shift={(90:\lab)}]center) {$y_{2}$};
        \node[font=\large] at ([shift={(270:\lab)}]center) {$y_{5}$};
        \draw[blue, dashed] (center) -- +(-60:\r);
        \node [above= \trt of center] {$A=a_{2}$};
        \coordinate (center) at (-1.5,0);
        \filldraw[draw=red,fill=red, fill opacity = 0.30] (center) -- +(0:\r) arc[start angle=0,end angle=300,radius=\r] -- cycle;
        \filldraw[draw=red,fill=red, fill opacity = 0.60] (center) -- +(300:\r) arc[start angle=300,end angle=360,radius=\r] -- cycle;
        \node[font=\large] at ([shift={(150:\lab)}]center) {$y_{3}$};
        \node[font=\large] at ([shift={(330:\lab)}]center) {$y_{6}$};
        \draw[red, dashed] (center) -- +(180:\r);

        \node [above= \trt of center] {$A=a_{3}$};
        \node at (-3, -2) {$P(Y^{a_{3}} > Y^{a_{2}}) = 2/3$};

        \draw[black, thick] (0, 2) -- (0, -2);
        
        \coordinate (center) at (1.5, 0);
        \filldraw[draw=blue,fill=blue, fill opacity = 0.20] (center) -- +(0:\r) arc[start angle=0,end angle=180,radius=\r] -- cycle;
        \filldraw[draw=blue,fill=blue, fill opacity = 0.50] (center) -- +(180:\r) arc[start angle=180,end angle=360,radius=\r] -- cycle;
        \node[font=\large] at ([shift={(90:\lab)}]center) {$y_{2}$};
        \node[font=\large] at ([shift={(270:\lab)}]center) {$y_{5}$};
        \draw[blue, dashed] (center) -- +(60:\r);
        \node [above= \trt of center] {$A=a_{2}$};
        \coordinate (center) at (4.5, 0);
        \filldraw[draw=red,fill=red, fill opacity = 0.30] (center) -- +(60:\r) arc[start angle=60,end angle=360,radius=\r] -- cycle;
        \filldraw[draw=red,fill=red, fill opacity = 0.60] (center) -- +(0:\r) arc[start angle=0,end angle=60,radius=\r] -- cycle;
        \node[font=\large] at ([shift={(210:\lab)}]center) {$y_{3}$};
        \node[font=\large] at ([shift={(30:\lab)}]center) {$y_{6}$};
        \draw[red, dashed] (center) -- +(180:\r);
        
        \node [above= \trt of center] {$A=a_{3}$};
        \node at (3, -2) {$P(Y^{a_{3}} > Y^{a_{2}}) = 1/2$};
    \end{tikzpicture}
    \caption{Maximum and minimum proportion of patients that can benefit from treatment $A=a_{3}$ over treatment $A=a_{2}$. As in Figure \ref{fig:rct 1}, dashed lines are intended as guides for visual comparison and do not indicate actual partitions of the data.}
    \label{fig:rct 2}
\end{figure*}
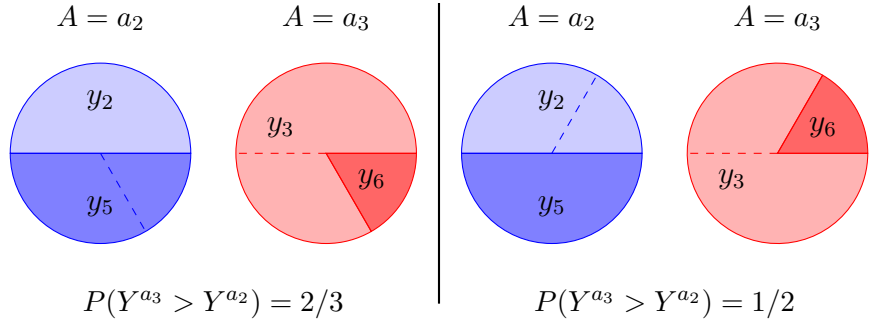

Figure \ref{fig:rct 2} shows the results of RCT-2 in graphical form. As in the previous example (RCT-1), bounds on the potential benefit can be calculated.

In the left set of Figure \ref{fig:rct 2}, all of the patients who would have developed XDR-TB under strong antibiotics, would only develop MDR-TB under the new antibiotics. In addition one-sixth of the total population would be fully cured without side-effects under the new antibiotics, as opposed to being fully cured with side-effects under the old antibiotics. Therefore, the maximum proportion of patients that would have benefited from the new treatment over treatment with the older antibiotics is $1/6+1/2=2/3$.

Again, as in RCT-1, the lower bound for the probability of benefit is $1/2$, as depicted in the right set of figures in Figure \ref{fig:rct 2}. Thus, in RCT-2, $P(\text{benefit}) = P(Y^{a_{3}} > Y^{a_{2}})$ is bounded as

\begin{equation}
    1/2\ \leq\ P(Y^{a_{3}} > Y^{a_{2}})\ \leq \ 2/3. 
\end{equation}

So, with at least probability half, the patient would benefit from taking the newer treatment $A=a_{3}$ over strong antibiotics, experiencing no side-effects. Except for strong determinisms in the population where $Y^{a_{2}} = y_{2}$ for every patient with $Y^{a_{3}} = y_{6}$, the probability of benefit is strictly bigger than $1/2$ (A rigorous argument is identical to the one for RCT-1 in Appendix \ref{app:frechet bounds}). The probability of counterfactual harm if the patient is prescribed the new treatment is indeed non-zero, but even if the patient is ``harmed'', the outcome of MDR-TB is still better than XDR-TB.

The expected utility under the new antibiotic treatment $\mathbb{E}[\mu_{1}(Y^{a_{3}})] = 3.5$ is also the same as for treatments $A=a_{1}$ or $A=a_{2}$ and there is no evidence of interventionist harm. So treatment with weak antibiotics is at least as good as treatment with strong antibiotics, given the objective of minimizing counterfactual harm.

\subsection{Interpreting the results of the two trials}
\label{sec:rct 3}

Data from RCT-1 and RCT-2 can be combined to directly compare the new antibiotic treatment $(A=a_{3})$ and no treatment $(A=a_{1})$:
\begin{align*}
    P(Y^{a_{3}} = y_{3}) &= P(Y = y_{3} \mid A=a_{3}) = 5/6, \\
    P(Y^{a_{3}} = y_{6}) &= P(Y = y_{6} \mid A=a_{3}) = 1/6, \\
    P(Y^{a_{1}} = y_{1}) &= P(Y = y_{1} \mid A=a_{1}) = 1/6, \\
    P(Y^{a_{1}} = y_{4}) &= P(Y = y_{4} \mid A=a_{1}) = 5/6.
\end{align*}

\begin{figure*}[htbp]
    \centering
    \begin{tikzpicture}
        \tikzmath{\r=1.2;\lab=0.7;\trt=1.5;}
        \coordinate (center) at (-4.5,0);
        \filldraw[draw=black,fill=black, fill opacity = 0.10] (center) -- +(60:\r) arc[start angle=60,end angle=120,radius=\r] -- cycle;
        \filldraw[draw=black,fill=black, fill opacity = 0.40] (center) -- +(120:\r) arc[start angle=120,end angle=420,radius=\r] -- cycle;
        \node[font=\large] at ([shift={(90:\lab)}]center) {$y_{1}$};
        \node[font=\large] at ([shift={(270:\lab)}]center) {$y_{4}$};
        \draw[black, dashed] (center) -- +(240:\r);
        \draw[black, dashed] (center) -- +(300:\r);
        \node [above= \trt of center] {$A=a_{1}$};
        \coordinate (center) at (-1.5,0);
        \filldraw[draw=red,fill=red, fill opacity = 0.30] (center) -- +(-60:\r) arc[start angle=-60,end angle=240,radius=\r] -- cycle;
        \filldraw[draw=red,fill=red, fill opacity = 0.60] (center) -- +(240:\r) arc[start angle=240,end angle=300,radius=\r] -- cycle;
        \node[font=\large] at ([shift={(90:\lab)}]center) {$y_{3}$};
        \node[font=\large] at ([shift={(270:\lab)}]center) {$y_{6}$};
        \draw[red, dashed] (center) -- +(60:\r);
        \draw[red, dashed] (center) -- +(120:\r);

        \node [above= \trt of center] {$A=a_{3}$};
        \node at (-3, -2) {$P(Y^{a_{3}} > Y^{a_{1}}) = 1/3$};

        \draw[black, thick] (0, 2) -- (0, -2);
        
        \coordinate (center) at (1.5, 0);
        \filldraw[draw=black,fill=black, fill opacity = 0.10] (center) -- +(60:\r) arc[start angle=60,end angle=120,radius=\r] -- cycle;
        \filldraw[draw=black,fill=black, fill opacity = 0.40] (center) -- +(120:\r) arc[start angle=120,end angle=420,radius=\r] -- cycle;
        \node[font=\large] at ([shift={(90:\lab)}]center) {$y_{1}$};
        \node[font=\large] at ([shift={(270:\lab)}]center) {$y_{4}$};
        \node [above= \trt of center] {$A=a_{1}$};
        \coordinate (center) at (4.5, 0);
        \filldraw[draw=red,fill=red, fill opacity = 0.30] (center) -- +(120:\r) arc[start angle=120,end angle=420,radius=\r] -- cycle;
        \filldraw[draw=red,fill=red, fill opacity = 0.60] (center) -- +(60:\r) arc[start angle=60,end angle=120,radius=\r] -- cycle;
        \node[font=\large] at ([shift={(90:\lab)}]center) {$y_{6}$};
        \node[font=\large] at ([shift={(270:\lab)}]center) {$y_{3}$};

        \node [above= \trt of center] {$A=a_{3}$};
        \node at (3, -2) {$P(Y^{a_{3}} > Y^{a_{1}}) = 1/6$};
    \end{tikzpicture}
    \caption{Maximum and minimum proportion of patients that can benefit from treatment $A=a_{3}$ over treatment $A=a_{1}$. Dashed lines serve purely as visual guides, same as Figures \ref{fig:rct 1} and \ref{fig:rct 2}.}
    \label{fig:rct 3}
\end{figure*}
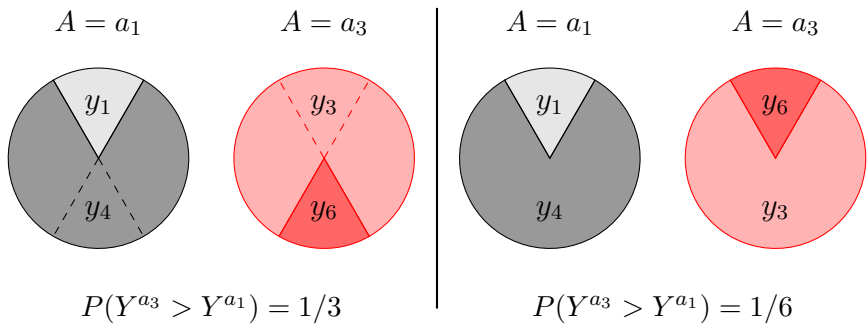

It is possible that the one-sixth of patients who would have died under no treatment are precisely the ones who get cured with the new antibiotics (rightmost pies in Figure \ref{fig:rct 3}). Thus, up to five-sixths of patients who would have latent TB under no treatment get MDR-TB after taking the new, weaker antibiotics. This argument is an intuitive justification for the following bounds,

\begin{equation}
    2/3\ \leq\ P(Y^{a_{1}} > Y^{a_{3}})\ \leq \ 5/6. 
\end{equation}

In summary, using the counterfactual definition, although strong antibiotics are as good as, or better than no treatment, and weak antibiotics are as good as, or better than strong antibiotics, weak antibiotics have a higher probability of potential harm compared to taking no treatment at all. This intransitivity may come across as paradoxical. In the two trials, corresponding to pair-wise comparisons, we minimized the probability of counterfactual harm yet we ended up with a treatment that is more harmful than the standard-of-care. %The doctor finds this conclusion confusing. At each step, he was minimizing the probability of counterfactual harm. So the result that giving no treatment $(A=a_{1})$ is less harmful than weak antibiotics $(A=a_{3})$ is difficult to justify.% In Section \ref{sec:discussion}, we resolve the apparent paradox and offer a solution.

% a majority of the population would be better off taking no treatment than taking weak antibiotics, which is a paradoxical conclusion. At each step, we were reducing the possibility of counterfactual harm against the standard-of-care. Intuitively, we would expect that going back to a previous standard-of-care increases counterfactual harm which is not the case when going back from weak antibiotics $(A=a_{3})$ to no treatment $(A=a_{1})$.

\section{Elaboration and explanation of the intransitivity}
\label{sec:discussion}
As we demonstrated, the objective of minimizing counterfactual harm is not one that guarantees a transitive ordering. One drawback of counterfactual harm is that the magnitude of difference between two outcomes is not taken into consideration, only whether a potential outcome is better or worse than the alternative. For example, in RCT-1, one-sixth of the patients benefit from taking treatment $A=a_{2}$, as they would have died if they were given no treatment. The magnitude of benefit in this case is fairly large.

In contrast, in RCT-2, one-sixth of the patients benefit from treatment $A=a_{3}$ by being fully cured without any side-effects over being fully cured with some side-effects under treatment $A=a_{2}$. Although the proportion of population that benefits is $1/6$ in both cases, the benefit that can be ascribed to patients in RCT-1 is inarguably greater than that for patients in RCT-2. 

Along the same line of thought, the patients in RCT-1 that do not benefit from treatment $A=a_{2}$ get XDR-TB as opposed to latent TB under no treatment. Both XDR-TB and latent TB can progress to active disease at a later stage. In RCT-2 the patients that do not benefit from treatment $A=a_{3}$ get MDR-TB as opposed to being fully cured under treatment $A=a_{2}$. The loss of benefit (or harm) ascribed to patients in RCT-1 is decidedly lower than that for patients in RCT-2.

Conceptually, with ordinal outcomes, we consider ``small'' benefits on an equal footing as ``large'' benefits, which leads to counter-intuitive conclusions and the intransitive ordering between treatments. Such problems do not occur when the outcomes are binary. In the binary case, there is no distinction between ``small'' or ``large'' benefits on the individual level: either an individual benefits or they do not. This is related to the fact that, in binary settings, interventionist decision rules coincide with rules based on the counterfactual criterion $P(\text{benefit}) \geq P(\text{harm})$ \citep{sarvet2023perspective, kallus2022s, ben2024policy}. However, when potential harm and benefit on binary outcomes are assigned different weights \citep{gelman2025russian, ben2024policy}, so-called asymmetric utility functions, the two notions of harm can give different decision rules \citep{mueller2023personalized, gelman2025russian, dawid2023personalised,sarvet2023perspective}.

As opposed to binary outcomes, with ordinal outcomes we can encode our understanding of ``small'' and ``large'' benefits by the utility function $\mu$. With the particular utility function $\mu_{1}$, the difference in utility between having latent TB and being fully cured with some side-effects was $1$ unit, which was the same as the difference in utility between being cured with side-effects v/s being cured without any side-effects. We can tailor the utility function to reflect our perceived magnitude of benefits. In some settings, ordinal outcomes are simply binarized to survival and death, or to progression-free-survival and not. This implicitly imposes a certain utility function.  

On the other hand, consider a utility function $\mu_{2}$ that maps to a number between $0$ and $10$, see Table \ref{tab:utility}. Death is the worst outcome, so it is assigned a utility of $0$. The various forms of TB are assigned utility between $3$ and $5$. Being fully cured with side-effects is not very different from being cured without side-effects, and both of these outcomes are much better than having TB in any form. Thus they are assigned utilities of $9$ and $10$ respectively. We can see that this utility function preserves the ordinal principle. Outcomes that are deemed better have higher utilities.

\begin{table}[htbp]
    \centering
    \caption{Sample utility function $\mu_{2}.$}
    \label{tab:utility}
    \begin{tabular}{c|c|c|c|c|c|c}
    \toprule
        Outcome & $y_{1}$ & $y_{2}$ & $y_{3}$ & $y_{4}$ & $y_{5}$ & $y_{6}$ \\
    \midrule
        Utility & 0 & 3 & 4 & 5 & 9 & 10 \\
    \bottomrule
    \end{tabular}
\end{table}

Based on our selected utility function $\mu_{2}$, for the three treatments $A=a_{1}, a_{2},$ and $a_{3}$, the expected utilities are $\mathbb{E}[\mu_{2}(Y^{a_{1}})] = 25/6$, $\mathbb{E}[\mu_{2}(Y^{a_{2}})] = 36/6$, and $\mathbb{E}[\mu_{2}(Y^{a_{3}})] = 30/6$. So we can say that `treatment with strong antibiotics' $(A=a_{2})$ is better than both `treatment with weak antibiotics' $(A=a_{3})$ and `no treatment' $(A=a_{1})$ at maximizing utility. 

However, this ordering is specific to our choice of utility function $\mu_{2}$. Consider another alternative utility function $\mu_{3}$ (Table \ref{tab:utility_3}) that also preserves the ordinal principle. The expected utilities are $\mathbb{E}[\mu_{3}(Y^{a_{1}})] = 40/6$, $\mathbb{E}[\mu_{3}(Y^{a_{2}})] = 30/6$, and $\mathbb{E}[\mu_{3}(Y^{a_{3}})] = 35/6$. The ordering of treatments with $\mu_{3}$ is the exact reverse of that achieved with $\mu_{2}$. It can be shown that any ordering between $a_{1}, a_{2},$ and $a_{3}$ can be achieved with an appropriate choice of utility function that still preserves the ordinal principle. 

The selection of an appropriate utility function has already been richly discussed in the literature, e.g., in the context of ``utility elicitation'' procedures \citep{chajewska1998utility, blythe2002visual, wang2003incremental}. For example, in the health economics literature Quality Adjusted Life Years (QALYs) are often used as a measure of utility \citep{torrance1976social, brazier1998deriving, drummond2015methods}. Contributing to this debate is beyond the scope of this article.

\begin{table}[htbp]
    \centering
    \caption{Sample utility function $\mu_{3}.$}
    \label{tab:utility_3}
    \begin{tabular}{c|c|c|c|c|c|c}
    \toprule
        Outcome & $y_{1}$ & $y_{2}$ & $y_{3}$ & $y_{4}$ & $y_{5}$ & $y_{6}$ \\
    \midrule
        Utility & 0 & 1 & 5 & 8 & 9 & 10 \\
    \bottomrule
    \end{tabular}
\end{table}

In some literature on counterfactual harm, the decision rule uses a weighted linear combination of benefit and harm \citep{ben2024policy, richens2022counterfactual, gelman2025russian}. The gain if an individual benefits is denoted by $u_{g}$, and the loss if an individual is harmed is $-u_{l}$, where $0 < u_{g} \leq u_{l}$ \citep{ben2024policy}. The greater loss in case the individual is harmed reinforces the principle of loss-aversion or ``do no harm''. Notably, this consideration of an asymmetric decision rule can result in interventionist and counterfactual rules differing in even binary settings \citep{ben2024policy, gelman2025russian}.

The weighted decision rule $u_{g}\cdot P(\text{benefit}) - u_{l}\cdot P(\text{harm}) \geq 0$, or equivalently $P(\text{benefit}) - w\cdot P(\text{harm}) \geq 0\ \text{where } w = u_{l}/u_{g}$, gives different treatment recommendations for different values of $w$. We apply such a rule to the examples of RCT-1 and RCT-2. For our particular examples, the probability of the outcome being exactly the same with both treatment options is zero, i.e. benefit and harm are complementary events with $P(\text{benefit}) + P(\text{harm}) = 1.$ As a result, the decision rule $P(\text{benefit}) - w\cdot P(\text{harm}) \geq 0$ can be rewritten as $P(\text{benefit}) \geq w/(1+w).$

In RCT-1 (Section \ref{sec:rct 1}) we compared `Strong Antibiotics' $(A=a_{2})$ to `No Treatment' $(A=a_{1})$ and obtained the bounds 
\begin{alignat*}{3}
    1/2 &\leq P(\text{benefit}; a_{2}, a_{1})\  &\leq 2/3,\\
    1/3 &\leq P(\text{harm}; a_{2}, a_{1}) &\leq 1/2.
\end{alignat*} 
If $w=1$, we could conclusively say that benefit was greater than (or equal to) harm. If we choose a value of $w \in (1, 2)$, it is possible that the expected counterfactual utility is positive with the given bounds, but we cannot conclusively say so. For $w > 2$, the expected counterfactual utility is always negative.

In RCT-2, we derived the same bounds when comparing `Weak Antibiotics' to `Strong Antibiotics'. For benefit and harm in RCT-2
\begin{alignat*}{3}
    1/2 &\leq P(\text{benefit}; a_{3}, a_{2})\ &\leq 2/3, \\
    1/3 &\leq P(\text{harm}; a_{3}, a_{2}) &\leq 1/2.
\end{alignat*}
Thus, the decision of which treatment among strong or weak antibiotics is the ``better'' one will depend on $w$ in the same manner as it did in RCT-1.

It is debatable whether the value of $w$ selected for RCT-1 should be the same as that for RCT-2. If we select the same value of $w$ for both comparisons, we are being consistent with our decision rule. However, it leads to the same problem that ``small'' harms in RCT-1 receive the same weight as ``big'' harms in RCT-2.

\section{Conclusion}
\label{sec:conclusion}

Applying counterfactual notions of harm holds the promise of improving decision making, e.g., by minimizing the fraction of population negatively affected by treatment \citep{kallus2022s, fava2024predicting, wu2024quantifying}. However, the existing work predominantly compares only two alternatives for treatment, with some exceptions \citep{richens2022counterfactual, beckers2022quantifying}. Due to this restriction, they never encounter the problem of intransitivity highlighted here.

Intransitivity and harm have been discussed conceptually in the philosophy literature \citep{norcross1998great, beckers2022quantifying} with qualitative arguments about comparing one treatment versus another. However, those works did not consider the definitions of counterfactual and interventionist harm (Definition \ref{def:counterfactual harm} and \ref{def:interventionist harm}).

Intransitivity can indicate that agents are being inconsistent or irrational in their choices and not objectively maximizing the desired outcome \citep{anand1995foundations}. Ordering treatments by counterfactual harm can lead to such intransitivity. On the other hand, the principle of minimizing interventionist harm produces a consistent transitive ordering once we have fixed a utility function. 

Defining counterfactual harm requires consideration of joint counterfactuals which are fundamentally unobservable. We can only place bounds on the joint counterfactual distribution with data from a perfectly executed randomized experiment. Studies that seek bounds tighter than the assumption-free bounds, e.g. \citet{ding2019decomposing, gechter2024generalizing, cui2023policy, wu2024quantifying}, employ assumptions on rank correlation between potential outcomes (individuals with higher outcomes under one treatment also have higher outcomes under the alternative treatment). Other studies on ordinal outcomes and the probability of benefit \citep{zhang2024identifying, de2025probability} make strong monotonicity assumptions. As \citet{gechter2024generalizing, zhang2024identifying, cui2023policy} described, these bounds quickly become uninformative if rank correlation or monotonicity assumptions are relaxed. Such assumptions are also fundamentally untestable, that is, they are cross-world assumptions \citep{richardson2013single}.

Bounds might be sharpened by leveraging baseline covariates, which, e.g., is a promising strategy in instrumental variable settings \cite{levis2025covariate}. Some success has been found in narrowing bounds when binary outcomes are considered; with increasing predictive power of the baseline covariates, the bounds on probability of benefit/harm might get tighter \citep{kallus2022s, mueller2023perspective, fava2024predicting}. However, the use of additional covariates does not guarantee point-identification of counterfactual harm. With non-binary (continuous) outcomes, we believe the improvement is less promising; the bounds remain uninformative even with covariate adjustment \citep{fava2024predicting, cui2021individualized}. Further, intransitivity in counterfactual harm is still a potential issue with non-binary outcomes.  % However, with continuous outcomes, the improvement is less promising; the bounds remain uninformative even with covariate adjustment \citep{fava2024predicting, cui2021individualized} without invoking additional cross-world assumptions.

The definition of interventionist harm does not concern joint counterfactuals, and thus does not require any additional assumptions on rank correlation. Interventionist harm is defined solely in terms of single-world expectations. Interventionist harm can thus be point-identified when population-level causal effects are point-identified. 

Lastly, we emphasize that our example was carefully constructed to exhibit intransitivity in the cleanest possible manner. The intransitivity holds for a wide class of joint distributions, including distributions satisfying rank correlation. In Appendix \ref{app:joint distributions}, we specify some instances of the full joint distribution for the TB example (Section \ref{sec:paradox}) under rank correlation, demonstrating that intransitivity fails to hold only if outcomes are perfectly negatively correlated. 

The conventional notion of counterfactual harm is inadequate as a decision criterion in non-binary settings. If algorithmic decision rules are to be used in practice, such as in clinical medicine, they must account for the possibility of multiple treatment or intervention options. Our results show that investigators should be careful when defining harm and their utility functions. 

\bibliographystyle{plainnat}
\bibliography{references}

@article{sarvet2023perspective,
  title={Perspective on ‘harm’ in personalized medicine},
  author={Sarvet, Aaron L and Stensrud, Mats J},
  journal={American Journal of Epidemiology},
  volume={194},
  number={6},
  pages={1743--1748},
  year={2025},
  publisher={Oxford University Press}
}

@article{levis2025covariate,
  title={Covariate-assisted bounds on causal effects with instrumental variables},
  author={Levis, Alexander W and Bonvini, Matteo and Zeng, Zhenghao and Keele, Luke and Kennedy, Edward H},
  journal={Journal of the Royal Statistical Society Series B: Statistical Methodology},
  pages={qkaf028},
  year={2025},
  publisher={Oxford University Press UK}
}

@article{mueller2023perspective,
  title={Perspective on ‘Harm’ in Personalized Medicine--An Alternative Perspective},
  author={Mueller, Scott and Pearl, Judea},
  journal={American Journal of Epidemiology},
  year={2023}
}

@article{sarvet2024rejoinder,
  title={Rejoinder to ``Perspectives on `harm' in personalized medicine--an alternative perspective''},
  author={Sarvet, Aaron L and Stensrud, Mats J},
  journal={American Journal of Epidemiology},
  volume={194},
  number={6},
  pages={1752--1755},
  year={2025},
  publisher={Oxford University Press}
}

@article{mueller2023personalized,
  title={Personalized decision making--A conceptual introduction},
  author={Mueller, Scott and Pearl, Judea},
  journal={Journal of Causal Inference},
  volume={11},
  number={1},
  pages={20220050},
  year={2023},
  publisher={De Gruyter}
}

@article{richens2022counterfactual,
  title={Counterfactual harm},
  author={Richens, Jonathan and Beard, Rory and Thompson, Daniel H},
  journal={Advances in Neural Information Processing Systems},
  volume={35},
  pages={36350--36365},
  year={2022}
}

@article{kallus2022s,
  title={What's the harm? sharp bounds on the fraction negatively affected by treatment},
  author={Kallus, Nathan},
  journal={Advances in Neural Information Processing Systems},
  volume={35},
  pages={15996--16009},
  year={2022}
}

@article{blyth1972some,
  title={Some probability paradoxes in choice from among random alternatives},
  author={Blyth, Colin R},
  journal={Journal of the American Statistical Association},
  volume={67},
  number={338},
  pages={366--373},
  year={1972},
  publisher={Taylor \& Francis}
}

@article{pasciuto2016mystery,
  title={The Mystery of the Non-Transitive Grime Dice},
  author={Pasciuto, Nicholas},
  journal={Undergraduate Review},
  volume={12},
  number={1},
  pages={107--115},
  year={2016}
}

@article{grime2017bizarre,
  title={The bizarre world of nontransitive dice: games for two or more players},
  author={Grime, James},
  journal={The College Mathematics Journal},
  volume={48},
  number={1},
  pages={2--9},
  year={2017},
  publisher={Taylor \& Francis}
}

@article{dawid2023personalised,
  title={Personalised decision-making without counterfactuals},
  author={Dawid, A Philip and Senn, Stephen},
  journal={arXiv preprint arXiv:2301.11976},
  year={2023}
}

@book{hernan2024causal,
  title={Causal Inference: What If},
  author={Hernan, M.A. and Robins, J.M.},
  isbn={9781420076165},
  lccn={2022050839},
  series={Chapman \& Hall/CRC Monographs on Statistics \& Applied Probab},
  year={2024},
  publisher={CRC Press}
}

@article{andrikyan2024artificial,
  title={Artificial intelligence-powered chatbots in search engines: a cross-sectional study on the quality and risks of drug information for patients},
  author={Andrikyan, Wahram and Sametinger, Sophie Marie and Kosfeld, Frithjof and Jung-Poppe, Lea and Fromm, Martin F and Maas, Renke and Nicolaus, Hagen F},
  journal={BMJ Quality \& Safety},
  year={2024},
  publisher={BMJ Publishing Group Ltd}
}

@inproceedings{beckers2022quantifying,
  title={Quantifying harm},
  author={Beckers, Sander and Chockler, Hana and Halpern, Joseph Y},
  booktitle = {Proceedings of the Thirty-Second International Joint Conference on
               Artificial Intelligence, {IJCAI-23}},
  publisher = {International Joint Conferences on Artificial Intelligence Organization},
  editor    = {Edith Elkind},
  pages     = {363--371},
  year      = {2023},
  month     = {8},
  note      = {Main Track},
  doi       = {10.24963/ijcai.2023/41},
  url       = {https://doi.org/10.24963/ijcai.2023/41},
}

@article{gechter2024generalizing,
  title={Generalizing the Results from Social Experiments: Theory and Evidence from India},
  author={Gechter, Michael},
  journal={Journal of Business \& Economic Statistics},
  volume={42},
  number={2},
  pages={801--811},
  year={2024},
  publisher={Taylor \& Francis}
}

@article{fan2010sharp,
  title={Sharp bounds on the distribution of treatment effects and their statistical inference},
  author={Fan, Yanqin and Park, Sang Soo},
  journal={Econometric Theory},
  volume={26},
  number={3},
  pages={931--951},
  year={2010},
  publisher={Cambridge University Press}
}

@article{ding2019decomposing,
  title={Decomposing treatment effect variation},
  author={Ding, Peng and Feller, Avi and Miratrix, Luke},
  journal={Journal of the American Statistical Association},
  volume={114},
  number={525},
  pages={304--317},
  year={2019},
  publisher={Taylor \& Francis}
}

@book{world2008implementing,
  title={Implementing the WHO Stop TB Strategy: a handbook for national TB control programmes},
  author={{World Health Organization}},
  year={2008},
  publisher={World Health Organization}
}

@book{world2010treatment,
  title={Treatment of tuberculosis: guidelines},
  author={{World Health Organization}},
  year={2010},
  publisher={World Health Organization}
}

@book{von1947theory,
  title={Theory of games and economic behavior, 2nd rev},
  author={Von Neumann, John and Morgenstern, Oskar},
  year={1947},
  publisher={Princeton university press}
}

@article{anand1995foundations,
  title={Foundations of rational choice under risk},
  author={Anand, Paul},
  journal={OUP Catalogue},
  year={1995},
  publisher={Oxford University Press}
}

@article{norcross1998great,
  title={Great harms from small benefits grow: how death can be outweighed by headaches},
  author={Norcross, Alastair},
  journal={Analysis},
  volume={58},
  number={2},
  pages={152--158},
  year={1998},
  publisher={JSTOR}
}

@article{cui2021individualized,
  title={Individualized decision-making under partial identification: Three perspectives, two optimality results, and one paradox},
  author={Cui, Yifan},
  journal={arXiv preprint arXiv:2110.10961},
  year={2021}
}

@article{stensrud2024optimal,
  title={Optimal regimes for algorithm-assisted human decision-making},
  author={Stensrud, Mats J and Laurendeau, Julien David and Sarvet, Aaron Leor},
  journal={Biometrika},
  volume={111},
  number={4},
  pages={1089--1108},
  year={2024},
  publisher={Oxford University Press}
}

@article{de2025probability,
  title={The Probability of Tiered Benefit: Partial Identification with Robust and Stable Inference},
  author={de Aguas, Johan and Krumscheid, Sebastian and Pensar, Johan and Biele, Guido},
  journal={arXiv preprint arXiv:2502.10049},
  year={2025}
}

@article{zhang2024identifying,
  title={Identifying and bounding the probability of necessity for causes of effects with ordinal outcomes},
  author={Zhang, Chao and Geng, Zhi and Li, Wei and Ding, Peng},
  journal={arXiv preprint arXiv:2411.01234},
  year={2024}
}

@article{richardson2013single,
  title={Single world intervention graphs (SWIGs): A unification of the counterfactual and graphical approaches to causality},
  author={Richardson, Thomas S and Robins, James M},
  journal={Center for the Statistics and the Social Sciences, University of Washington Series. Working Paper},
  volume={128},
  number={30},
  pages={2013},
  year={2013},
  publisher={Citeseer}
}

@article{frechet1935generalisation,
  title={G{\'e}n{\'e}ralisation du th{\'e}oreme des probabilit{\'e}s totales},
  author={Fr{\'e}chet, Maurice},
  journal={Fundamenta mathematicae},
  volume={25},
  number={1},
  pages={379--387},
  year={1935},
  publisher={Polska Akademia Nauk. Instytut Matematyczny PAN}
}

@incollection{ruschendorf1991frechet,
  title={Fr{\'e}chet-bounds and their applications},
  author={R{\"u}schendorf, Ludger},
  booktitle={Advances in Probability Distributions with Given Marginals: beyond the copulas},
  pages={151--187},
  year={1991},
  publisher={Springer}
}

@inproceedings{wang2003incremental,
  title={Incremental utility elicitation with the minimax regret decision criterion},
  author={Wang, Tianhan and Boutilier, Craig},
  booktitle={Ijcai},
  volume={3},
  pages={309--316},
  year={2003}
}

@inproceedings{blythe2002visual,
  title={Visual exploration and incremental utility elicitation},
  author={Blythe, Jim},
  booktitle={AAAI/IAAI},
  pages={526--532},
  year={2002}
}

@inproceedings{chajewska1998utility,
  title={Utility Elicitation as a Classification Problem.},
  author={Chajewska, Urszula and Getoor, Lise and Norman, Joseph and Shahar, Yuval},
  booktitle={UAI},
  volume={98},
  pages={79--88},
  year={1998}
}

@article{torrance1976social,
  title={Social preferences for health states: an empirical evaluation of three measurement techniques},
  author={Torrance, George W},
  journal={Socio-economic planning sciences},
  volume={10},
  number={3},
  pages={129--136},
  year={1976},
  publisher={Elsevier}
}

@book{drummond2015methods,
  title={Methods for the economic evaluation of health care programmes},
  author={Drummond, Michael F and Sculpher, Mark J and Claxton, Karl and Stoddart, Greg L and Torrance, George W},
  year={2015},
  publisher={Oxford university press}
}

@article{brazier1998deriving,
  title={Deriving a preference-based single index from the UK SF-36 Health Survey},
  author={Brazier, John and Usherwood, Tim and Harper, Rosemary and Thomas, Kate},
  journal={Journal of clinical epidemiology},
  volume={51},
  number={11},
  pages={1115--1128},
  year={1998},
  publisher={Elsevier}
}

@article{fava2024predicting,
  title={Predicting the Distribution of Treatment Effects: A Covariate-Adjustment Approach},
  author={Fava, Bruno},
  journal={arXiv preprint arXiv:2407.14635},
  year={2024}
}

@article{cui2023policy,
  title={Policy learning with distributional welfare},
  author={Cui, Yifan and Han, Sukjin},
  journal={arXiv preprint arXiv:2311.15878},
  year={2023}
}

@article{ben2024policy,
  title={Policy learning with asymmetric counterfactual utilities},
  author={Ben-Michael, Eli and Imai, Kosuke and Jiang, Zhichao},
  journal={Journal of the American Statistical Association},
  volume={119},
  number={548},
  pages={3045--3058},
  year={2024},
  publisher={Taylor \& Francis}
}

@article{gelman2025russian,
  title={Russian roulette: the need for stochastic potential outcomes when utilities depend on counterfactuals},
  author={Gelman, Andrew and Mikhaeil, Jonas M},
  journal={Biometrika},
  volume={112},
  number={4},
  pages={asaf062},
  year={2025},
  publisher={Oxford University Press}
}

@article{wu2024quantifying,
  title={Quantifying individual risk for binary outcome},
  author={Wu, Peng and Ding, Peng and Geng, Zhi and Liu, Yue},
  journal={arXiv preprint arXiv:2402.10537},
  year={2024}
}

\section*{Appendix}
\label{sec:appendix}

\renewcommand{\thesubsection}{\Alph{subsection}}

\subsection{Mathematical derivation of bounds from RCT-1}
\label{app:frechet bounds}

To derive bounds on $P(Y^{a_{2}} > Y^{a_{1}})$ consider the following four principal stratum probabilities:
\begin{align*}
    P(Y^{a_{1}} = y_{1},\ Y^{a_{2}} = y_{2}) &= p_{1}, \\
    P(Y^{a_{1}} = y_{1},\ Y^{a_{2}} = y_{5}) &= p_{2}, \\
    P(Y^{a_{1}} = y_{4},\ Y^{a_{2}} = y_{2}) &= p_{3}, \\
    P(Y^{a_{1}} = y_{4},\ Y^{a_{2}} = y_{5}) &= p_{4}.
\end{align*}
We know that $p_{1},\ p_{2},\ p_{3},\ p_{4} \geq 0$ and $p_{1} + p_{2} + p_{3} + p_{4} = 1$. Additionally, from the observed marginals in RCT-1, we can impose the following constraints, 
\begin{alignat*}{3}
    P(Y^{a_{1}} = y_{1}) &= p_{1} + p_{2} = 1/6 &&\Rightarrow p_{1} \leq 1/6, \\
    P(Y^{a_{2}} = y_{2}) &= p_{1} + p_{3} = 1/2 &&\Rightarrow p_{3} = 1/2 - p_{1}, \\
    P(Y^{a_{2}} = y_{5}) &= p_{2} + p_{4} = 1/2.
    \intertext{The probability of benefit is given by}
    P(Y^{a_{2}} > Y^{a_{1}}) &= p_{1} + p_{2} + p_{4} = 1/2 + p_{1},\qquad &&(0\leq p_{1} \leq 1/6)\\
    \intertext{and respectively, the probability of counterfactual harm,}
    P(Y^{a_{2}} < Y^{a_{1}}) &= p_{3} = 1/2 - p_{1}.  &&(0\leq p_{1} \leq 1/6)
\end{alignat*}
These expressions give us the bounds from RCT-1 (Figure \ref{fig:rct 1}). Furthermore, $P(Y^{a_{2}} > Y^{a_{1}}) = P(Y^{a_{2}} < Y^{a_{1}}) = 1/2$ occurs only if $p_{1} = 0$, i.e. there are no patients belonging to the principal stratum $(Y^{a_{1}} = y_{1},\ Y^{a_{2}} = y_{2})$. Thus, strong antibiotics are strictly more beneficial than harmful compared to no treatment, except if there are strong determinisms in the data.

The bounds from RCT-2 can be derived in an identical manner. In RCT-2, the edge case of $P(Y^{a_{3}} > Y^{a_{2}}) = P(Y^{a_{3}} < Y^{a_{2}}) = 1/2$ occurs only if $P(Y^{a_{2}} = y_{5},\ Y^{a_{3}} = y_{6}) = 0$, otherwise weak antibiotics are strictly more beneficial than harmful compared to strong antibiotics.

\subsection{Rank correlated joint distributions}
\label{app:joint distributions}

In the Conclusion section, we commented that the intransitive order between TB treatments occurs for a wide class of joint distributions, including distributions satisfying rank correlation. We expand on the statement in this Appendix.

First we describe rank correlation conceptually. The term ``rank correlation'' is used to indicate that the joint potential outcomes are correlated. Suppose that an individual has an outcome higher than the population average under treatment $A=a_{1}$, thereby having a higher rank amongst all individuals in the population. Conceptually, if this individual's outcome under treatment $A=a_{2}$ would be expected to be higher than the population average under treatment $A=a_{2}$, then we say that the joint potential outcomes are positively rank correlated. Rank correlation can indicate the plausible assumption that healthier individuals are more likely to survive under no treatment, and are also more likely to respond to treatment and thus get cured. So joint outcomes are positively correlated.% Rank correlation can thus also be interpreted in terms of frailty scores. There is an unmeasured variable $U$ that is informative of potential outcomes $Y^{a}$ \citep{}.

To fix ideas, consider an individual with the best possible outcome under no treatment who can have a better-than-average, but not the best, outcome under treatment. Such a scenario still represents positive rank correlation, but not exact correlation.

The assumption of strong positive rank correlation is sometimes invoked to sharpen bounds on potential benefit/harm \citep{gechter2024generalizing, ding2019decomposing, cui2021individualized, wu2024quantifying}. For instance, \citet{ding2019decomposing} commented that the maximum of potential harm occurs when joint outcomes are perfectly negatively associated, which is unlikely to happen in practice. Similarly, \citet{wu2024quantifying} commented that a patient's unmeasured health status can affect potential outcomes under both treatment and control, making the joint outcomes positively correlated. Therefore, bounds on potential harm can be sharpened by assuming joint outcomes are positively rank correlated.

In Section \ref{sec:paradox}, we described only the marginal distributions of outcomes, without making any additional assumptions on the joint distribution of potential outcomes under different treatments. We now exactly specify two joint distributions of the example from Section \ref{sec:paradox}. First, we corroborate \citet{ding2019decomposing}, showing that the the edge case where $P(\text{benefit}) = P(\text{harm}) = 1/2$ is attained precisely when joint outcomes are perfectly negatively correlated. 

In RCT-1, we compared outcomes under no treatment $(A=a_{1})$ and under strong antibiotics $(A=a_{2})$. Consider the one-sixth of individuals that die of TB $(Y^{a_{1}} = y_{1})$. Such individuals are ranked the lowest in the population in terms of outcomes under no treatment. If joint outcomes are negatively rank correlated, these individuals tend to be cured with side-effects, the best possible outcome when given strong antibiotics $(Y^{a_{2}} = y_{5})$.

The joint potential outcomes follow the law $P(Y^{a_{1}} = y_{1}, Y^{a_{2}} = y_{5}) = 1/6,\ P(Y^{a_{1}} = y_{1}, Y^{a_{2}} = y_{2}) = 0$, which is depicted in the right set of pies in Figure \ref{fig:rct 1}. Such a distribution corresponds to the maximum potential harm $P(Y^{a_{2}} > Y^{a_{1}}) = 1/2$.

Similarly, in the right set of pies in Figure \ref{fig:rct 2}, the upper bound for harm, $P(Y^{a_{3}} > Y^{a_{2}}) = 1/2$, is attained only if outcomes satisfy the law $P(Y^{a_{2}} = y_{2}, Y^{a_{3}} = y_{6}) = 1/6,\ P(Y^{a_{2}} = y_{5}, Y^{a_{3}} = y_{6}) = 0$. In this scenario, the individuals with the best possible outcome under weak antibiotics ($y_{6}:$ Cured without side effects) always have the worst possible outcome under strong antibiotics ($y_{2}:$ XDR-TB), meaning outcomes are negatively rank correlated.

Negative rank correlation is theoretically possible, but has been argued to be implausible \citep{ding2019decomposing, wu2024quantifying}. Thus, the equality $P(\text{benefit}) = P(\text{harm}) = 1/2$ only happens in an arguably contrived special case.

Next, we specify the joint distribution with positive rank correlation. Take the most healthy individuals who have the best possible outcomes under each treatment. For these individuals the joint outcomes are $P(Y^{a_{1}}= y_{4}, Y^{a_{2}}= y_{5}, Y^{a_{3}}= y_{6}) = 1/6$. Next, the slightly less healthier individuals do not get cured with weak antibiotics, but do respond to strong antibiotics $P(Y^{a_{1}}= y_{4}, Y^{a_{2}}= y_{5}, Y^{a_{3}}= y_{3}) = 1/3$. The other individuals are divided as $P(Y^{a_{1}}= y_{4}, Y^{a_{2}}= y_{2}, Y^{a_{3}}= y_{6}) = 1/3$ and $P(Y^{a_{1}}= y_{1}, Y^{a_{2}}= y_{2}, Y^{a_{3}}= y_{3}) = 1/6$. Thus, the weakest individuals are the ones that would die under no treatment, get XDR-TB under strong antibiotics, and get MDR-TB when treated with weak antibiotics.

Given this joint distribution of outcomes, we have that $P(Y^{a_{2}} > Y^{a_{1}}) = P(Y^{a_{3}} > Y^{a_{2}}) = P(Y^{a_{1}} > Y^{a_{3}}) = 2/3$ and the treatments are maximally intransitive. We conclude that the assumption of positive rank correlation, which has been invoked to sharpen bounds on potential benefit/harm \citep{wu2024quantifying, gechter2024generalizing}, does not guarantee transitivity of counterfactual harm comparisons. In fact, even with perfect knowledge of the joint counterfactual distribution, we cannot discount the possibility of intransitivity.

\end{document}